\documentstyle[12pt]{article}
\begin{document}

\centerline{\large\bf Robustness of the inflationary perturbation
spectrum}
\centerline{\large\bf to trans-Planckian physics}

\medskip

\centerline{\rm A.A. Starobinsky}

\medskip

\centerline{\it Landau Institute for Theoretical Physics,
Russian Academy of Sciences,}
\centerline{\it 117334 Moscow, Russia}
\centerline{and}
\centerline{\it Research Center for the Early Universe,
The University of Tokyo,}
\centerline{\it 113-0033 Tokyo, Japan}

\vspace{0.5cm}

\centerline{\bf Abstract}

\medskip
\noindent

It is investigated if predictions of the inflationary scenario
regarding spectra of scalar and tensor perturbations generated from
quantum vacuum fluctuations are robust with respect to a
modification of the dispersion law for frequencies beyond the Planck
scale. For a large class of such modifications of special and general
relativity, for which the WKB condition is not violated at super-high
frequencies, the predictions remain unchanged. The opposite
possibility is excluded by the absence of large amount of created
particles due to the present Universe expansion. Creation of
particles in the quantum state minimizing the energy density of a
given mode at the moment of Planck boundary crossing is prohibited
by the latter argument, too (contrary to creation in the adiabatic
vacuum state which is very small now).

\medskip


\vspace{0.5cm}

Approximately flat spectrum of scalar and tensor perturbations
generated from quantum vacuum fluctuations at a quasi-de Sitter
(inflationary) state in the early Universe is certainly the most
important prediction of the inflationary scenario since it can be
directly tested and falsified using observational data. Fortunately,
all existing and constantly accumulating data, instead of falsifying,
confirm these predictions (within observational errors). The other
observational prediction of the simplest variants of the inflationary
scenario -- the approximate flatness of the Universe, $|\Omega_{tot}
-1|\ll 1$ -- is actually a consequence of the first one since an
isotropic part of the spatial curvature may be considered as a
monopole perturbation with respect to the spatially flat
Friedmann-Robertson-Walker (FRW) background. Note that the first
quantitatively correct derivation of perturbation spectra {\em after}
inflation was first obtained in~\cite{St79} in the case of tensor
perturbations (gravitational waves) and in~\cite{HSG82} in the case
of scalar (adiabatic) perturbations. For completeness, one should
mention two important intermediate steps on the way to the right
answer for scalar perturbations made between 1979 and 1982:
in~\cite {L80}, the first estimate of scalar perturbations after
inflation was made according to which scalar and tensor
perturbations are of the same order of magnitude, while
in~\cite{M81}, the spectrum of scalar perturbations {\em during}
inflation was calculated for the Starobinsky inflationary
model~\cite{St80} (however, the actual amplitude of scalar
perturbations {\em after} inflation was still significantly
underestimated in both these papers).

Therefore, it is very important to investigate the validity of
assumptions on which this prediction is based
\footnote {Results of the present paper partially intersect with
those obtained in the recent papers~\cite{T00,NP01} (which appeared
when this paper was prepared for publication) and are in general
agreement with them whenever intersect.}$^)$. All derivations of
perturbation spectra use quantum field theory in classical curved
space-time or semiclassical quantum cosmology. Both these approaches
are valid and essentially equivalent if $H\ll M_P$ where $H\equiv
\dot a/a$,~$a(t)$ is the scale factor of a flat FRW cosmological
model, the dot denotes time derivative, $M_P=\sqrt G$, and
$\hbar=c=1$ is put throughout the paper. On the other hand,
comparison of the predicted spectrum with observational data shows
that $H$ should be less than $\sim 10^{-5}M_P$ at least during last
$70$ e-folds of inflation. So, the assumption $H\ll M_P$ is
{\em required} and {\em self-consistent}, if we are speaking about
inflationary models having relation to reality. Recently it was
questioned if inflationary predictions are robust with respect to a
change in the so-called "trans-Planckian physics". What is meant by
this term is some {\it ad hoc} modification of special and general
relativity leading to violation of the Lorentz invariance and to
deviation of the dispersion law $\omega (k)$ for field quanta from
the linear one for frequencies (energies) $\omega>M_P$, where $k$ is
a particle wave number (momentum). In the absence of the Lorentz
invariance, a preferred system of reference appears (in which this
dispersion law is written). Usually, it is identified with the basic
cosmological system of reference which is at rest with respect to
spatially averaged matter in the Universe.

Initially, "trans-Planckian physics" was introduced to obtain a new
way of derivation of the Hawking radiation from black holes. In this
case, it was shown that the spectrum of the Hawking radiation does
not depend on a concrete form of the dispersion law $\omega(k)$ at
$k\to \infty$~\cite{U95,BMPS95,CJ96}. On the other hand, an opposite
result was recently claimed in \cite{MB00} regarding the inflationary
perturbation spectrum. There exists no self-consistent theory of such
a modification leading to some unique dispersion law $\omega(k)$, but
arguments showing that this possibility should not be considered as
logically impossible are based either on higher-dimensional models of
the Universe (see, e.g., the recent paper \cite{CKR00}), or on
condensed matter analogs of gravity \cite{V00,V01} which do not have
too much symmetry at the most fundamental level. So, $\omega(k)$
should be considered as some fixed but unknown function at the
present state-of-the-art.

The very possibility of trans-Planckian physics affecting the
(supposedly known) sub-Planckian one is due to the expansion of the
Universe. This expansion gradually shifts all modes of quantum fields
from the former region to the latter one. Really, for a FRW model
with the metric
\begin{equation}
ds^2=dt^2-a^2(t)\, dl^2
\end{equation}
where $dl^2$ is the 3D Euclidean space interval (spatial curvature
may be always neglected), spatial dependence of a given mode of a
quantum field may
be taken as $\exp(in_{\mu}x^{\mu}),~\mu=1,2,3$. Then the frequency
$\omega=n/a(t),~n=|{\bf n}|=const$ in the ultra-relativistic (but
still Lorentzian) limit. This red-shifting occurs equally well in the
early and the present-day Universe. So, any effect connected with
trans-Planckian physics can be observed now, too; inflation (i.e.,
the epoch when $|\dot H|\ll H^2$) is not specific for that at all.

We will model metric fluctuations by a massless, minimally coupled
scalar field satisfying the equation $\nabla_i\nabla^i\phi =0$. This
form is sufficient for both scalar perturbations for which the
effective mass satisfies the condition $|m^2|\ll H^2$ necessary for
inflation, and for tensor perturbations since their amplitude
satisfies the same wave equation in the FRW Universe filled by any
matter with no non-diagonal pressure perturbations
($\delta p_{\mu\nu}\propto\delta_{\mu\nu}$). Also it is assumed that
$H\ll M_P$. Then the equation for the time-dependent part of $\phi_n$
reads
\begin{equation}
\ddot\phi_n + 3H\dot\phi_n +\omega^2\left({n\over a}\right)\phi_n = 0
\end{equation}
with $\omega(k)=k$ for $\omega\ll M_P$. Solutions of this equation
have the WKB form for $H\ll \omega \ll M_P$:
\begin{equation}
\phi_n = {\alpha_n\over \sqrt{2n} a}e^{-in\eta}+{\beta_n\over
\sqrt{2n}a}e^{in\eta}~,~~\eta=\int{dt\over a(t)}
\label{WKB}
\end{equation}
where $\alpha_n,~\beta_n=const$ and $|\alpha_n|^2-|\beta_n|^2=1$ for
any quantum state, if the quantum field $\hat \phi$ is second
quantized and $\phi_n\exp\left(in_{\mu}x^{\mu}\right)(2\pi)^{-3/2}$
is the c-number coefficient of the Fock annihilation operator
$\hat a_{\bf n}$. The average number of created pairs is $N(n)=
|\beta_n|^2$. Therefore, whatever the trans-Planckian physics is
(namely, what is the form of $\omega(k)$ and what is the initial
condition for $\phi_n$ for $t\to -\infty$), once $\omega\ll M_P$, we
may say that the field mode~(\ref{WKB}) emerges from the Planck
boundary $n=M_Pa$ in some quantum state characterized by $\alpha_n$
and $\beta_n$. In particular, the rate of growth of the average
energy density of particles with $\omega\ll M_P$ is
\begin{equation}
{d(<\varepsilon> a^4)\over a^4\,dt}={gM^4H\over 2\pi^2}N(n)|_{n=Ma}
\label{grow}
\end{equation}
where $g=1$ for scalars and $g=2$ for gravitons. $M$ is an
auxiliary mass satisfying $H\ll M < M_P$ for which $\omega(M)=M$
with sufficient accuracy (for estimates, we will take $M=M_P$).
It follows from time translation invariance that $N^{(0)}(n)$ is
independent on $n$. Here $N^{(0)}$ means the part of $N(n)$ which
does not depend on background space-time curvature at the moment of
Planck boundary crossing ($n=M_Pa$).

Let us first consider the case when the WKB condition for $\phi_n$
is satisfied for all $n\gg Ha$ including the trans-Planckian
region $n>M_Pa$. Then the natural and self-consistent choice of
the initial condition for $\phi_n$ is the adiabatic vacuum for
$t\to -\infty$:
\begin{equation}
\phi_n={1\over \sqrt{2\omega_n a^3}}\exp(-i\int\omega_n dt)~.
\label{in}
\end{equation}
Note that this mode is not in the minimum energy density state at
finite $t$, in particular, at the Planck boundary crossing (we
return to the discussion of this point below). Eq.~(\ref{in})
reduces to Eq.~(\ref{WKB}) with $\beta_n=0,~\alpha_n=1$ in the
sub-Planckian region. Then it just coincides with the initial
condition for $\phi_n$ used in the standard calculation of the
spectrum of inflationary perturbations. Thus, no correction to the
standard result arises in this case irrespective of the form of
$\omega(n/a)$.

The necessary condition for the WKB behaviour is $|\dot \omega|
\ll \omega^2$, or
\begin{equation}
{H|d(1/\omega(k))|\over d\ln k}\ll 1~,~~k=n/a
\end{equation}
for all $k>M_P$. Since $H/M_P$ is already a small parameter and
$\omega(k)$ presumably does not depend on $H$ for $k\gg H$, this
inequality is satisfied practically always,
if $\omega$ does not become zero either for $k\to\infty$ or at some
finite $k_0>M_P$ (another dangerous case is when $d\omega/dk$
diverges at a finite $k=k_0$, in particular, if $\omega\propto
(k_0-k)^{\gamma}$ with $-1<\gamma<0$ or $\omega \approx \omega_0
+ \omega_1(k_0-k)^{\gamma},~0<\gamma<1$). As a consequence,
$N^{(0)}=0$ for the dispersion law
$\omega(k)=M\tanh^{1/m}[(k/M)^{m}],~m>0$ proposed by
Unruh~\cite{U95}, for $\omega^2=k^2[1+b_m(k/M)^{2m}]$ with positive
$m$ and $b_m$ considered in~\cite{CJ96,MB00}, and for the dependence
$\omega^2 = [M\ln (1+k/M)]^2$ introduced in~\cite{Pad98} (see also
\cite{K01}).

Still, there exist exceptional forms of $\omega(k)$ for which the WKB
behaviour is not valid for some $k>M_P$. In particular, this refers
to the case $\omega^2=k^2[1+b_m(k/M)^{2m}]$ with $b_m<0$ and to the
dispersion law introduced in the recent paper~\cite{MBK01} for which
$\omega(k)\to 0$ at $k\to\infty$. {\it A priori}, such a possibility
may not be excluded. Then there is {\em no} preferred initial
condition for $\phi_n$, and it is not possible to define a unique
initial vacuum state. So, in this case $N^{(0)}\not= 0$ generically,
i.e., creation of pairs in the expanding Universe occurs due to
trans-Planckian physics.

However, nature tells us that such an effect is infinitesimally
small, if exists at all. Really, from the evident condition that
created ultra-relativistic particles do not significantly
contribute to the present energy density in the Universe, it
follows that $N^{(0)}\stackrel{<}{\sim} H_0^2/M_P^2\sim 10^{-122}$
where $H_0=H(t=t_0)$ is the Hubble constant. Thus, curvature
independent particle creation in the expanding Universe due to
trans-Planckian physics is very strongly suppressed in any case
because of observational data. Of course, the corresponding change
in the inflationary perturbation spectrum is negligible, too
(the relative correction is $\sim |\beta_n|=\sqrt{N^{(0)}}$).

Finally, let us consider a more subtle effect: creation of particles
due to both trans-Planckian physics and background space-time
curvature in the expanding Universe. Then $N(n)\sim H^2/M_P^2$
where $H$ is estimated at the moment of high energy boundary crossing
$n=Ma(t)$. Certainly, corrections to the inflationary spectrum are
already negligible ($\sim H/M_P< 10^{-5}$) in this case.
Nevertheless, even such a small effect can be significantly
restricted. An example of this effect arises if we would assume that
modes crossing the boundary $n=Ma$ are in the exactly minimum energy
density state just at this moment, i.e., $\dot\phi_n=-in\phi_n/a=
-i\sqrt{n/2}\,a^{-2}$ and $\varepsilon_n\equiv (|\dot\phi_n|^2+n^2
a^{-2}|\phi_n|^2)/2= n/2a^4$ for each mode at the moment $t=t_n$ when
$n=Ma$. On the other hand, the adiabatic vacuum for each mode has the
larger energy density
\begin{equation}
\varepsilon_n={n\over 2a^4}\left(1+{H^2a^2\over 2n^2}\right)
\end{equation}
(see, e.g., \cite{ZS71,FPH74}). Note that this excess is due to
vacuum polarization only. Of course, this assumption may be
immediately criticized from the logical point of view since such
a state ceases to diagonalize the mode Hamiltonian and minimize its
energy density for all other moments of time $t\not= t_n$.
Nevertheless, let us consider its implications.

Writing as, e.g., in \cite{ZS71}:
\begin{equation}
\phi_n(t) = (2\omega_na^3)^{-1/2}\left(\alpha_n(t)\exp (-i\int
\omega_n\,dt)+ \beta_n(t)\exp (i\int\omega_n\, dt)\right)~,
\end{equation}
\begin{equation}
\dot\phi_n(t)=-i\left({\omega_n\over 2a^3}\right)^{1/2}\left(
\alpha_n(t)\exp (-i\int\omega_n\, dt)-\beta_n(t)\exp (i\int
\omega_n\, dt)\right)~,
\end{equation}
so that $\alpha_n(t_n)=1,~\beta_n(t_n)=0$ for the Heisenberg quantum
state of each mode $|\psi_n>$ which minimizes its Hamiltonian and
energy density at the moment $t=t_n$ when $n=Ma$, we obtain the
following system of equations for $\alpha_n(t)$ and $\beta_n(t)$:
\begin{eqnarray}
\dot\alpha_n={1\over 2}\left({\dot \omega\over\omega}+3{\dot a\over
a}\right)e^{2i\int\omega_n dt}\beta_n~, \\
\dot\beta_n={1\over 2}\left({\dot \omega\over\omega}+3{\dot a\over
a}\right)e^{-2i\int\omega_n dt}\alpha_n
\end{eqnarray}
with the additional condition $|\alpha_n|^2-|\beta_n|^2=1$. If
$\omega \gg H$, $\beta_n$ is small and $\alpha_n\approx 1$. For
$t\ge t_n$, one may take $\omega_n\approx n/a$. Then $\beta_n=
-{iH(t_n)\over 2M}\exp (-2i\eta(t_n))$ plus a strongly oscillating
term. So,
\begin{equation}
N(n)=|\beta_n(\infty)|^2= {H^2(t_n)\over 4M^2}~.
\label{N1}
\end{equation}
If the cosmological constant is neglected and the present law of the
Universe expansion is taken as $a(t)\propto t^{2/3}$, then
$N(n)\propto n^{-3}$ for particle energies close to $M_P$ at the
present time. Integrating Eq.~(\ref{grow}) with $N(n)$ from
Eq.~(\ref{N1}), we obtain $\varepsilon_g=M^2/9\pi^2t^2$ for
gravitons. For $M\sim M_P$, $\varepsilon_g \sim H^2/G$
that contradicts the assumption that $a(t)\propto t^{2/3}$. In
other words, this model of particle creation by trans-Planckian
physics results in a significant part of the present total energy
density of matter in the Universe being contained in gravitons
with energies $\sim M_P$ that is not compatible with the observed
behaviour of $a(t)$. Similar arguments show that there may be no
term
\begin{equation}
N(n)=N^{(1)}(n){|R(t_n)|\over M_P^2}
\end{equation}
with $N^{(1)}\sim 1$ in Eq.~(\ref{grow}). Here $R$ is the scalar
curvature.

On the other hand, for the adiabatic vacuum state in the WKB
regime, the quantity $\beta_n = iH(t)a(t)\exp (-2in\eta)/2n$ in the
leading order, so it approaches zero for $t\to\infty$. Note that
creation of real gravitons does occur in the next order ($N=N^{(2)}
(n)R^2/M_P^2$), and even without any violation of the Lorentz
invariance~\cite{ZS77}. In the latter case, the effect is due to
violation of the WKB approximation at ultra-low, not ultra-high,
frequencies $\omega \sim H$. Also, the notion of "vacuum" as the
state of a minimum energy density may be restored in the following
non-rigorous sense: the adiabatic vacuum of each mode $n$ in the WKB
regime has the lowest energy density compared to other quantum
states, if the energy density is averaged ("coarse grained") over a
time interval $\Delta t \gg \omega_n^{-1}$ in accordance with the
energy uncertainty relation.

So, whatever occurs in the trans-Planckian region,
observational evidence show that creation of particles due to mode
transition from the trans-Planckian region to the sub-Planckian
one is absent with a very high accuracy. Standard predictions about
perturbations generated during inflation are not altered by this
hypothetical mechanism, too.

The author is thankful to Profs. K. Sato and M. Kawasaki for
hospitality in RESCEU, the University of Tokyo. This research was
also partially supported by the Russian Fund for Fundamental
Research, grants 99-02-16224 and 00-15-96699.

\vfill
\end{document}